\documentclass[sigconf]{acmart}
\usepackage[nolist]{acronym}
\usepackage{graphicx}
\usepackage{subcaption}
\usepackage{wrapfig}
\usepackage{hyperref}
\AtBeginDocument{%
  \providecommand\BibTeX{{%
    \normalfont B\kern-0.5em{\scshape i\kern-0.25em b}\kern-0.8em\TeX}}}

\copyrightyear{2022} 
\acmYear{2022} 
\setcopyright{acmlicensed}\acmConference[AutomotiveUI '22]{14th International Conference on Automotive User Interfaces and Interactive Vehicular Applications}{September 17--20, 2022}{Seoul, Republic of Korea}
\acmBooktitle{14th International Conference on Automotive User Interfaces and Interactive Vehicular Applications (AutomotiveUI '22), September 17--20, 2022, Seoul, Republic of Korea}
\acmPrice{}
\acmDOI{10.1145/3543174.3545173}
\acmISBN{978-1-4503-9415-4/22/09}

\begin{document}

\begin{acronym}[]
	\acro{AI}{Artificial Intelligence}
	\acro{ACC}{Adaptive Cruise Control}
	\acro{ADAS}{Advanced Driver Assistance System}
	\acro{CAN}{Controller Area Network}
	\acro{ECU}{Electronic Control Unit}
	\acro{GDPR}{General Data Protection Regulation}
	\acro{GOMS}{Goals, Operators, Methods, Selection rules}
	\acro{HCI}{Human-Computer Interaction}
	\acro{HMI}{Human-Machine Interaction}
	\acro{HU}{Head Unit}
	\acro{IVIS}{In-Vehicle Information System}
	\acro{KLM}{Keystroke-Level Model}
	\acro{KPI}{Key Performance Indicator}
	\acro{LKA}{Lane Keeping Assist}
	\acro{LCA}{Lane Centering Assist}
	\acro{MGD}{Mean Glance Duration}
	\acro{OEM}{Original Equipment Manufacturer}
	\acro{OTA}{Over-The-Air}
	\acro{AOI}{Area of Interest}
	\acro{RF}{Random Forest}
	\acro{SA}{Steering Assist}
	\acro{SHAP}{SHapley Additive exPlanation}
	\acro{TGD}{Total Glance Duration}
	\acro{UCD}{User-centered Design}
	\acro{UX}{User Experience}
\end{acronym}

\title[The Effect of Driving Automation on Touchscreen Interactions and Glance Behavior]{How Do Drivers Self-Regulate their Secondary Task Engagements? The Effect of Driving Automation on Touchscreen Interactions and Glance Behavior}




\author{Patrick Ebel}
\email{ebel@cs.uni-koeln.de}
\orcid{0000-0002-4437-2821}
\affiliation{%
	\institution{University of Cologne}
	\city{Cologne}
	\country{Germany}}

\author{Moritz Berger}
\email{moritz.berger@imbie.uni-bonn.de}
\orcid{0000-0002-0656-5286}
\affiliation{%
	\institution{University of Bonn}
	\city{Bonn}
	\country{Germany}}

\author{Christoph Lingenfelder}
\email{christoph.lingenfelder@mercedes-benz.com}
\orcid{0000-0001-9417-5116}
\affiliation{%
	\institution{MBition}
	\city{Berlin}
	\country{Germany}}

\author{Andreas Vogelsang}
\email{vogelsang@cs.uni-koeln.de}
\orcid{0000-0003-1041-0815}
\affiliation{%
	\institution{University of Cologne}
	\city{Cologne}
	\country{Germany}}

\renewcommand{\shortauthors}{Ebel et al.}

\begin{abstract}
With ever-improving driver assistance systems and large touchscreens becoming the main in-vehicle interface, drivers are more tempted than ever to engage in distracting non-driving-related tasks. However, little research exists on how driving automation affects drivers’ self-regulation when interacting with center stack touchscreens. To investigate this, we employ multilevel models on a real-world driving dataset consisting of 10,139 sequences. Our results show significant differences in drivers' interaction and glance behavior in response to varying levels of driving automation, vehicle speed, and road curvature. During partially automated driving, drivers are not only more likely to engage in secondary touchscreen tasks, but their mean glance duration toward the touchscreen also increases by 12\,\% (Level 1) and 20\,\% (Level 2) compared to manual driving. We further show that the effect of driving automation on drivers' self-regulation is larger than that of vehicle speed and road curvature. The derived knowledge can facilitate the safety evaluation of infotainment systems and the development of context-aware driver monitoring systems.

\end{abstract}



\maketitle

\section{Introduction}

Driver distraction is one of the main causes of motor vehicle crashes. The primary goal of automated driving functions like \ac{ACC} and \ac{LCA}, apart from making driving more comfortable, is to make driving safer. Multiple studies show that these systems can make driving safer by an increased time headway and that they reduce the incidence of critical situations~\cite{faber.2012, ervin.2005}. However, even though automated driving functions are more widely available and powerful than ever, the number of crashes based on human error due to distraction stagnated in recent years~\cite{NHTSA.2021}. Studies show that driving automation does not only positively affect driving safety but also tends to increase the margins in which drivers consider it safe to engage in non-driving-related tasks~\cite{risteska.2021, morando.2019, dunn.2020}. To interact with \acp{IVIS} or mobile phones while driving, drivers need to distribute their attention between the primary driving task and the non-driving-related secondary task. Although drivers are proven to self-regulate their secondary task engagements based on driving demands~\cite{christoph.2019, onate-vega.2020, oviedo-trespalacios.2019}, this task-switching behavior is directly associated with an increased crash risk~\cite{dingus.2016}. This is particularly critical as drivers tend to overestimate the capabilities of automated driving functions~\cite{deguzman.2021} potentially making it more likely to engage in non-driving-related tasks~\cite{dunn.2020} in situations in which they are supposed to monitor these functions constantly~\cite{SAE.2021}. 

As modern \acp{IVIS} continue to improve and large center stack touchscreens are becoming the main interface between driver and vehicle, the temptation for drivers to interact with them is likely to increase~\cite{starkey.2020}. A deep understanding of how drivers self-regulate their secondary task engagements in response to varying driving demands can facilitate the design of \acp{IVIS} that are safe to use in all situations~\cite{ebel.2021, ebel.2021b}. Knowing what naturally feels safe for drivers can also, improve attention management systems to provide situation-dependent interventions when inappropriate self-regulation is detected~\cite{risteska.2021}. 

To better understand how drivers adapt their engagement in secondary touchscreen tasks, we investigate the effect of driving automation (manual vs. \ac{ACC} vs. \ac{ACC} + \ac{LCA}), vehicle speed, and road curvature on drivers' tactical and operational self-regulation. We further show how the effect of driving automation depends on vehicle speed and road curvature. Therefore, we employ multilevel modeling on a real-world driving dataset consisting of 10,139 user interaction sequences and the accompanying driving and eye tracking data. 

To evaluate tactical self-regulation, we fit generalized linear mixed models estimating the probability of drivers interacting with specific UI elements. Our results show that drivers self-regulate their interaction behavior differently across the UI elements. During ACC+LCA driving, the odds of a driver interacting with a map element are, for example, 2.3 times as high as for manual driving. The probability to interact with a regular button, however, remains similar.

Furthermore, we measure drivers' operational self-regulation as glance behavior adaptions. The multilevel modeling results indicate that drivers adapt their glance behavior based on automation level, vehicle speed, and road curvature. Across all driving situations, the mean glance duration increases by 12\,\% for ACC driving compared to manual driving and by 20\,\% for ACC+LCA driving. The odds that drivers perform a glance longer than 2 seconds are 2.7 and 3.6 times as high, respectively.


\subsection{Driver Distraction and Self-Regulation}\label{ch:DriverDistraction}
During distracted driving, drivers divide their attention between the primary driving task and a non-driving-related secondary task like interacting with the \ac{IVIS}~\cite{regan.2009}. To maintain safe driving and to mitigate the risks associated with the secondary task demands, drivers self-regulate their multitasking behavior~\cite{rudin-brown.2013a}. According to~\citeauthor{rudin-brown.2013a}, self-regulation can be intentional or unintentional and occurs at three distinct levels derived from Michon's driver task model~\cite{michon.1985}: strategic, tactical, and operational~\cite{rudin-brown.2013a}. 

Strategic self-regulation describes driver decisions that are made on a timescale of minutes or more~\cite{rudin-brown.2013a} and are often constant over a trip. \citeauthor{young.2010}, for example, report that some drivers state that they never engage in a secondary task in heavy traffic, in poor weather conditions, or when driving at nighttime~\cite{young.2010}. \citeauthor{oviedo-trespalacios.2019} modeled strategic self-regulation as the decision to pull over to perform a secondary task~\cite{oviedo-trespalacios.2019}. In this case, drivers made the strategic decision to not engage in secondary tasks while driving. 

Tactical self-regulation refers to a driver's decision to engage or not engage in a specific task in a certain situation. Tactical decisions are made in the time frame of seconds~\cite{rudin-brown.2013a} and are continuously updated while driving. Many studies investigate the engagement in mobile phone tasks while driving and have shown that drivers are less likely to engage in a visual manual phone task when driving demands are high (high speed, sharp turns, etc.)~\cite{tivesten.2015, hancox.2013, oviedo-trespalacios.2018a, oviedo-trespalacios.2019, ismaeel.2020}. \citeauthor{tivesten.2015} show that drivers use information about the upcoming driving demand to decide whether or not to engage in a secondary task~\cite{tivesten.2015}. 
Somewhat contrary results are presented by \citeauthor{horrey.2009}, who found that although drivers were well aware of the demands of specific traffic situations, it had little influence on performing the secondary task~\cite{horrey.2009}. This is consistent with findings presented by~\citeauthor{carsten.2017} who show that drivers stopped engaging in easy tasks when the driving demand increased but continued to engage in more demanding secondary tasks~\cite{carsten.2017}. \citeauthor{liang.2015} found that even though drivers avoided initiating a secondary task before an immediate transition to higher driving demands, they did also not postpone their secondary task engagement when driving demand was already high~\cite{liang.2015}. \citeauthor{carsten.2017} and~\citeauthor{liang.2015} argue that more work is needed to evaluate the factors influencing tactical self-regulation.

Operational self-regulation describes behavioral adaptions made by the driver while actively engaging in a secondary task. This implies that on the strategic and tactical level the driver already decided to engage in a secondary task. Operational self-regulation can be bidirectional. On the one hand, research has shown that drivers adjust their driving behavior in terms of vehicle speed, lane position, or time headway~\cite{schneidereit.2017, morgenstern.2020, onate-vega.2020, oviedo-trespalacios.2018,choudhary.2017}. On the other hand, recent findings indicate that drivers also adjust their secondary task engagement in response to variations in driving demand. \citeauthor{oviedo-trespalacios.2019} found that drivers temporarily stopped the use of mobile phones to cope with varying driving demands~\cite{oviedo-trespalacios.2019}. Similarly, in a test track experiment, \citeauthor{liang.2015} show that drivers adjust their time-sharing behavior according to driving demands~\cite{liang.2015}. In addition, \citeauthor{tivesten.2014} state that drivers allow for more distraction in less demanding situations. In a naturalistic driving study, drivers performed shorter off-road glances during turning when a lead vehicle was present and when oncoming traffic was detected~\cite{tivesten.2014}. \citeauthor{tivesten.2014} further argue that drivers prioritize secondary tasks over monitoring the driving environment, especially in low-speed situations. Accordingly, \citeauthor{risteska.2021} show that drivers' off-path glances decrease in situations with higher visual difficulty~\cite{risteska.2021}.

\subsection{The Effect of Driving Automation on Self-Regulation}

Multiple studies have investigated the effect of partially automated driving (Level 1 and Level 2 according to SAE J3016 ~\cite{SAE.2021}) on drivers' secondary task engagement. As laid out in the following, studies suggest that more automation may result in less driver engagement and, therefore, a lower capability to correctly assess the current driving situation.

\citeauthor{lin.2019} investigate drivers' self-regulation in Level 2 driving according to the levels of situation awareness as proposed by~\citeauthor{schomig.2013}. On the control level, which corresponds to operational self-regulation, they found that drivers tend to pause their engagement in case of urgent hazards but continue to engage with a more frequent task switching behavior in case of less urgent hazards.

In addition, many studies investigated how drivers allocate their visual attention while driving with partial automation activated. Results from the Virginia Connected Corridors Level 2 naturalistic driving study~\cite{dunn.2020} indicate that the use of Level 2 automation (i.e., \ac{ACC}+\ac{LCA}) led to drivers spending less time with their eyes on driving-related tasks. In accordance,~\citeauthor{gaspar.2019} found that with partial automation activated, drivers made longer single off-road glances and had longer maximum total-eyes-off-road times~\cite{gaspar.2019}. This finding is complemented by the results presented by~\citeauthor{yang.2021} who also found that off-road glances were longer in automated driving conditions and additionally investigated the effect of different levels of distraction. They found that off-road glances were longer for highly distracting secondary tasks~\cite{yang.2021}.
\citeauthor{noble.2021} assessed the effect of \ac{ACC}, \ac{LCA}, and \ac{ACC} + \ac{LCA} on drivers' glance behavior and secondary task engagement. They found that during ACC+LCA driving, drivers execute longer and more frequent glances away from the road~\cite{noble.2021}. They, however, did not find significant differences in the mean off-road glance duration nor in the tactical self-regulation when \ac{ACC} + \ac{LCA} was active. Another naturalistic driving study is presented by~\citeauthor{morando.2019} who found a significant decrease in the percentage of time with eyes on the road center when using \ac{ACC} + \ac{LKA}~\cite{morando.2019}. In a subsequent study, the authors investigated drivers' glance behavior during disengagements of Tesla's Autopilot in naturalistic highway driving~\cite{morando.2021}. Whereas they found that all off-road glances tended to be longer with AP compared to manual driving, the difference was particularly big for glances down and toward the center stack. The mean glance duration increased by 0.3 seconds and the proportion of glances longer than 2 seconds increased by 425\% in Autopilot conditions compared to manual driving.

\subsection{Research Questions}\label{ch:RQs}
We identify two main research gaps in the current state of the art: (1) Current work is mainly focused on self-regulation when interacting with mobile phones or when engaging in general secondary tasks such as eating, drinking, or talking to a passenger. No work addresses operational and tactical self-regulatory behavior during explicit interactions with \acp{IVIS}. (2) Whereas multiple studies investigate the general effect of partial automation on drivers' self-regulation, there is yet no detailed investigation on the interdependencies between driving automation, vehicle speed, and road curvature.

Considering that modern \acp{IVIS} are increasingly complex and incorporate nearly all the functionality of smartphones and that \ac{ACC} and \ac{LCA} are becoming more capable and accessible, we argue that both aspects need to be examined in more detail. Therefore, we aim to answer the following research questions:
	\begin{itemize}
		\item[\textbf{RQ1:}] To what extent do drivers self-regulate their behavior on the tactical level when engaging in secondary touchscreen tasks depending on driving automation, vehicle speed, and road curvature?
		\item[\textbf{RQ2:}] To what extent do drivers self-regulate their behavior on the operational level when engaging in secondary touchscreen tasks depending on driving automation, vehicle speed, and road curvature?
		\item[\textbf{RQ3:}] Does the effect of driving automation on drivers' operational self-regulation vary in response to different driving situations?
	\end{itemize}
	


\section{Method}

\subsection{Data Source and Data Collection}
The presented results are based on 10,139 interaction sequences extracted from 2,755 individual trips generated by more than 100 individual test vehicles from mid-October 2021 to mid-January 2022. The vehicles are part of the internal test fleet of Mercedes-Benz and are used for a variety of testing procedures but also for transfer and leisure drives of employees. All vehicles that are equipped with the most recent software architecture, a stereo camera for glance detection, and ACC and LCA technology contributed to the data collection. ACC automates the longitudinal control and LCA supports the lateral control keeping the car in the center of the lane. Both systems work at speeds between 0\,km/h and 210\,km/h.

The data used in this study was collected over the air, leveraging the data collection and processing framework of Mercedes-Benz. A more detailed explanation is given in previous work by \citeauthor{ebel.2021a}~\cite{ebel.2021a}. We analyze touchscreen interactions, driving data (vehicle speed, steering wheel angle, and level of driving automation), and eye tracking data.
Steering wheel angle and vehicle speed are logged at a frequency of $4\,Hz$. For each user interaction on the center stack touchscreen a data point that consists of a timestamp, and the interactive UI element is logged.
To control for confounding factors, we also collect the seat belt signal of all passenger seats and an additional UI-related signal that allows us to monitor all changes in the UI that may occur due to inputs via modalities other than the touchscreen.

Glance data is acquired using a stereo camera located in the instrument cluster behind the steering wheel. The eye tracking is primarily based on the pupil-corneal reflection technique~\cite{merchant.1967}, which is used in the majority of remote eye tracking devices~\cite{hutchinson.1989}. The driver’s field of view is divided into different \acp{AOI} and the system continuously keeps track of the driver’s gaze by mapping it to one of the \acp{AOI}. The true positive rate of the \acp{AOI} describing the center stack touchscreen is above 90 percent. The system used in this research is a production system without the ability to capture raw video data.

\subsection{Data Processing}

\subsubsection{User Interaction Data}
In contrast to controlled experiments, there is no predefined secondary task that the drivers have to perform. We are not aware of drivers' intentions nor of interactions that semantically belong to the same task. We rather observe drivers' natural behavior in an unbiased setting. We, therefore, extract user interaction sequences based on the assumption that drivers disengaged from the secondary task when they do not interact with the touchscreen for more than 10 seconds. The next interaction is then considered the starting point of a new interaction sequence. 
Further, we discard all sequences in which UI changes occur that we cannot map to touch interactions. These changes are either caused by user interactions via other modalities or by some system-initiated notification.

\subsubsection{Eye Tracking Data}
To improve the quality of the eye tracking data, we apply multiple filtering steps, partially adapted from related work~\cite{morando.2019} and in accordance with ISO 15007-1:2020~\cite{ISO15007}. First, we extract all glances toward the center stack touchscreen between the first and the last interaction of each interaction sequence. To handle periods of tracking loss shorter than $300\,ms$, we interpolate gaps in which the preceding \ac{AOI} is equal to the succeeding one. Furthermore, all glances shorter than $120\,ms$, which is considered the shortest fixation that humans can control~\cite{ISO15007}, are interpolated. Accordingly, we also interpolate losses of tracking shorter than $120\,ms$. Finally, to remove blinks we interpolate all eyelid closures shorter than $500\, ms$.

\subsubsection{Driving Data}
The driving data consists of vehicle speed, steering wheel angle, and automation level. First, we extract all data that is relevant for a specific interaction sequence. For each sequence, we consider the vehicle speed and steering wheel angle data from two seconds before the first interaction until 2 seconds after the last interaction. This allows us to compute more stable aggregate statistics for very short sequences. We discard all sequences for which deviations in the logging frequency or sensor outages were detected.

\subsubsection{Final Filtering and Data Description}\label{ch:filtering}

\begin{figure}
	\centering
	\includegraphics[width = 0.7\linewidth]{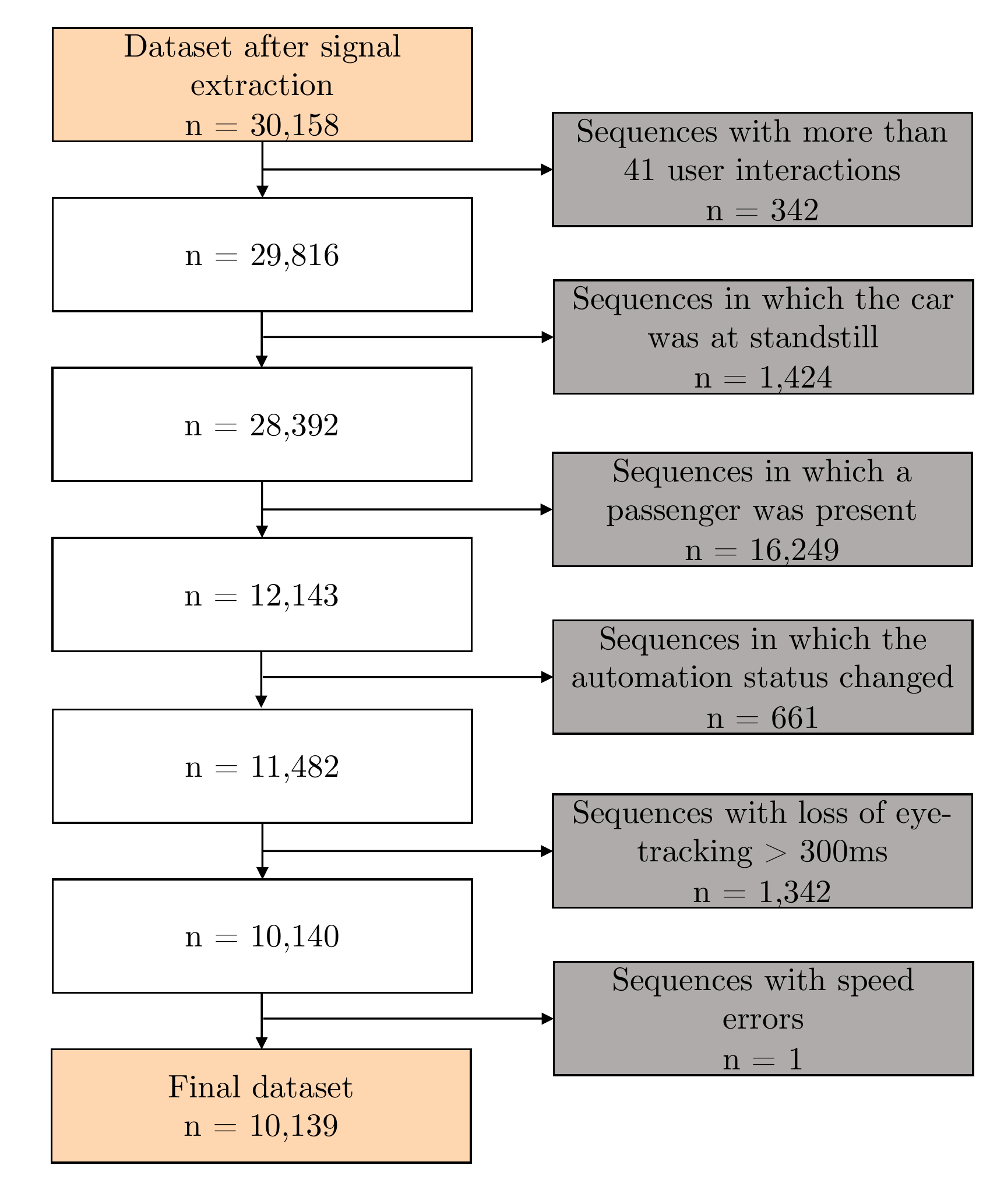} 
	\caption{Data filtering procedure} 
	\label{fig:DataExclusion} 
\end{figure}

After individual signal extraction, the dataset contains 30,158 sequences. To improve data quality and control for confounding factors, we apply strict exclusion criteria as visualized in Figure~\ref{fig:DataExclusion}. We discard all sequences with more than 41 interactions, which corresponds to the $99^{th}$ percentile of the distribution of interactions per sequence. We further discard all sequences in which the car was stationary as we are only interested in self-regulation while driving. To control for potential distractions or interactions by a passenger, we delete all sequences in which a passenger seat belt buckle was latched. Additionally, we discard all sequences during which the automation level changed because it is not clear to which automation level we can assign these sequences. This includes sequences in which the driver selected another automation level, overwrote the current level by accelerating or braking, or in which the automation was deactivated due to external factors like a loss of lane marking or bad weather conditions. Lastly, all sequences with a loss of eye tracking larger than $300\,ms$ and sequences with errors in the speed or steering wheel angle signal are discarded. The final dataset contains 10,139 sequences of which 8,655 are manual driving, 770 are ACC driving, and 714 are ACC+LCA driving\footnote{The dataset statistics are given here: \url{https://doi.org/10.6084/m9.figshare.19668918.v1}}.

\subsection{Statistical Modeling}
As stated in Section~\ref{ch:RQs}, we investigate how drivers' tactical and operational self-regulation changes in response to different levels of driving automation and driving contexts. In the following, we introduce the dependent and independent variables, and the statistical models we use. We define statistical significance at the level of $\alpha=0.05$.

\subsubsection{Dependent Variables}
We analyzed the following dependent variables:

\textbf{UI Interactions.} Current approaches are mostly investigating tactical self-regulation by comparing the likelihood of a driver engaging in a specific secondary task given different driving situations. We aim to investigate drivers' tactical self-regulation in greater detail, enabling us to draw conclusions about the UI design itself. Therefore, we choose driver interactions with specific UI elements (dichotomous) as a dependent variable. The different categories of UI elements are listed in Table~\ref{tab:UIelements}.

\begin{table*}
	\small
	\caption{Overview of the different UI elements.}
	\label{tab:UIelements}
	\centering
		\begin{tabular}{@{}ll@{}}
		    \toprule
			Category & Description \\
			\toprule
			Button         & General buttons like push buttons or radio buttons\\
			List           & List containers used, for example, to present destination suggestions\\
			Homebar        & Static homebar on the bottom of all screens (contains home button, music and climate controls) \\
			AppIcon        & Application icons on the home screen, used to start an application\\
			Tab         & Tab bar used to navigate between different views or subtasks\\
			Map       & Map viewer that displays a map and allows for interactions with it \\
			Keyboard       & Virtual keyboard or number pad to enter text \\
            CoverFlow      & Animated widget that, for example, allows flipping through album covers \\
			Slider         & Vertical or horizontal sliders used, for example, when changing the volume \\
			RemoteUI       & Apple CarPlay or AndroidAuto \\
			ControlBar     & Menu controls to show context menus or popups \\
			ClickGuard     & Non-interactable background elements\\
			Other          & UI elements that do not fit any of the above categories \\
			Unknown        & UI elements for which the identifier is not specified \\
			\bottomrule
		\end{tabular}
\end{table*}

\textbf{Mean Glance Duration.} The \textit{mean glance duration} is a continuous variable computed as the sum of all individual glance durations toward the center stack touchscreen during a sequence divided by the total number of glances per sequence. 

\textbf{Long Glance.} The dichotomous variable \textit{long glance} indicates whether a driver glanced at the center stack touchscreen for more than two seconds. Eyes-off-road glances longer than two seconds are associated with an increased crash risk~\cite{Klauer.2006}. The proportion of such long glances is an important factor in evaluating drivers' operational self-regulation.

\subsubsection{Independent Variables}
The dependent variables are analyzed with respect to the following independent variables:

\textbf{Automation Level.} The automation level is a categorical variable with three distinct levels: \textit{manual, ACC, and ACC+LCA}. According to SAE J3016~\cite{SAE.2021}, these levels correspond to Level 0, 1, and 2 of driving automation. The automation level is constant throughout each sequence.

\textbf{Vehicle Speed.} The vehicle speed is a categorical variable with three levels: $0\,km/h < v \leq 50\,km/h, 50\,km/h < v \leq 100\,km/h\,, v > 100\,km/h$. It is computed as the average speed across a sequence. 

\textbf{Road Curvature.} The road curvature is a categorical variable with two levels: \textit{straight} or \textit{curved}. An interaction sequence is classified as curved if the maximum absolute steering wheel angle is greater than 50\textdegree or if the absolute average steering wheel angle is greater than 5\textdegree.

\subsubsection{Models}

As our data structure is hierarchical (interaction sequences are nested within trips) and our study design unbalanced (not all combinations of the independent variables are observed at all trips), we employ mixed-effect models also referred to as multilevel models~\cite{hox.1998}. Multilevel models are well suited for unbalances designs and account for grouping hierarchies~\cite{magezi.2015}, making them a suitable solution to test our hypotheses.

We performed all our analyses using R Statistical Software (v4.1.3) \cite{rcoreteam.2022}. We used the \textit{lme4} package (v.1.1.28)~\cite{bates.2015} to build the multilevel models, obtained p-values via the \textit{lmertest} package (v.3.1.3)~\cite{kuznetsova.2017}, and computed the pairwise post-hoc tests using the \textit{emmmeans} package (v.1.7.3)~\cite{lenth.2022}. Regression tables were generated using the \textit{stargazer} package (v.5.2.3)~\cite{hlavac.2022}.

\textbf{User Interaction Models.}
We evaluate tactical self-regulation by modeling the probability of drivers to interact with a specific UI element. To estimate the probability of a driver to engage with these elements, we fit one logistic mixed-effect model with random intercepts for each UI element. We include \textit{automation level}, \textit{vehicle speed}, and \textit{road curvature} as fixed effects and the trip during which the sequence was recorded as random effect. We omit interaction effects because none of the two-way or three-way interactions between the independent variables proved to be significant.

\textbf{Glance Behavior Models.}
To estimate the \textit{mean glance duration}, we fit two linear mixed-effect models with random intercepts. An exploratory data analysis showed that the distribution of the mean glance duration is heavily right-skewed. To satisfy the model assumption of normally distributed residuals we, therefore, apply a log transformation.
In Model 1 we estimate the effect of driving automation on the mean glance duration across all driving situations by only selecting the automation level as fixed effect. 
In Model 2 we add the vehicle speed and road curvature as additional fixed effects and allow for interaction effects. To account for the hierarchical structure of our data we include the trip during which an interaction sequence was recorded as random effect for both models.

To estimate drivers' long glance behavior, we fit two logistic mixed effect models with random intercepts. In Model 3 we select the automation level as fixed effect and in Model 4 we add the vehicle speed and road curvature as fixed effects and model all interactions between the independent variables. The trip information is, again, entered as random effect.

Visual inspection of residual plots and Q-Q plots of the final models did not reveal any obvious deviations from homoscedasticity or normality. We use Satterthwaite's degrees of freedom approximation to obtain p-values and evaluate significances~\cite{luke.2017}. For post-hoc pairwise comparisons we use Tukey's multiple comparison method~\cite{tukey.1949}.

\section{Results}
In the following, we present the results obtained by fitting the above-introduced models to the 10,139 interaction sequences.  

\subsection{Tactical Self-Regulation}
Table~\ref{tab:InteractionResults} shows the parameters of the user interaction models for all UI elements that occur in more than 10\,\% of all sequences.\footnote{The results of the other models are provided here: \url{https://doi.org/10.6084/m9.figshare.19668918.v1}}. The results suggest that drivers adapt their interaction behavior with the center stack touchscreen based on automation status, vehicle speed, and road curvature.

\begin{table*}
\centering 
\small
  \caption{Generalized linear mixed-effect models describing the probability of the driver interacting with Tab, List, Button, Homebar, or AppIcon UI elements during an interaction sequence. For each model, the intercept and the coefficients describing the effect of the independent variables are given along with the estimated standard error. Coefficients and standard errors correspond to log odds ratios.} 
  \label{tab:InteractionResults} 
\begin{tabular}{@{\extracolsep{5pt}}lrrrrrr} 
\\[-1.8ex]\hline 
\hline \\[-1.8ex] 
 & \multicolumn{6}{c}{\textit{Dependent variable:}} \\ 
\cline{2-7} 
\\[-1.8ex] & Tab & List & Map & Button & Homebar & AppIcon \\ 
\hline \\[-1.8ex] 
 Intercept & $-$1.50$^{***}$ (0.06) & $-$1.64$^{***}$ (0.06) & $-$2.86$^{***}$ (0.11) & $-$0.83$^{***}$ (0.05) & $-$0.53$^{***}$ (0.05) & $-$2.14$^{***}$ (0.07) \\ 
  ACC & 0.51$^{***}$ (0.10) & 0.54$^{***}$ (0.11) & 0.79$^{***}$ (0.14) & 0.32$^{***}$ (0.09) & 0.37$^{***}$ (0.09) & 0.54$^{***}$ (0.11) \\ 
  ACC+LCA & 0.36$^{***}$ (0.11) & 0.49$^{***}$ (0.11) & 0.83$^{***}$ (0.14) & $-$0.01 (0.10) & 0.18 (0.09) & 0.36$^{**}$ (0.12) \\ 
  50--100 & $-$0.05 (0.07) & 0.15$^{*}$ (0.07) & 0.07 (0.09) & 0.20$^{***}$ (0.06) & 0.19$^{***}$ (0.05) & 0.30$^{***}$ (0.07) \\ 
  100+ & $-$0.16 (0.09) & 0.04 (0.09) & 0.06 (0.11) & $-$0.08 (0.07) & $-$0.01 (0.07) & 0.25$^{**}$ (0.10) \\ 
  curved & $-$0.26$^{***}$ (0.07) & $-$0.24$^{***}$ (0.07) & $-$0.38$^{***}$ (0.09) & $-$0.12$^{*}$ (0.05) & 0.002 (0.05) & $-$0.21$^{**}$ (0.07) \\ 
 \hline \\[-1.8ex] 
Akaike Inf. Crit. & 9,772.59 & 10,045.28 & 7,058.58 & 12,719.40 & 13,276.97 & 8,369.28 \\ 
Bayesian Inf. Crit. & 9,823.16 & 10,095.85 & 7,109.15 & 12,769.97 & 13,327.54 & 8,419.85 \\ 
\hline 
\hline \\[-1.8ex] 
\textit{Note:}  & \multicolumn{6}{r}{$^{*}$p$<$0.05; $^{**}$p$<$0.01; $^{***}$p$<$0.001} \\ 
\end{tabular} 
\end{table*} 

The $\beta$ coefficients for the independent variables given in Table~\ref{tab:InteractionResults} represent log-odds ratios. This means that, keeping everything else constant, a change in the predictor by one level results in a $e^{\beta}$ increase or decrease in the odds that the driver interacts with the respective UI element. Considering the \textit{Map} model the coefficients can be interpreted as follows: During ACC driving the odds that a driver performs a map interaction are $e^{0.79} \approx 2.2$ times as high as the odds of performing the same interaction in manual driving. In contrast, when driving in curved conditions, the odds for a map interaction are $e^{-0.38} \approx 0.68$ the odds of performing a map interaction in straight driving.

\begin{figure*}%
    \centering
    \begin{subfigure}{0.47\linewidth}
    \includegraphics[width=\linewidth]{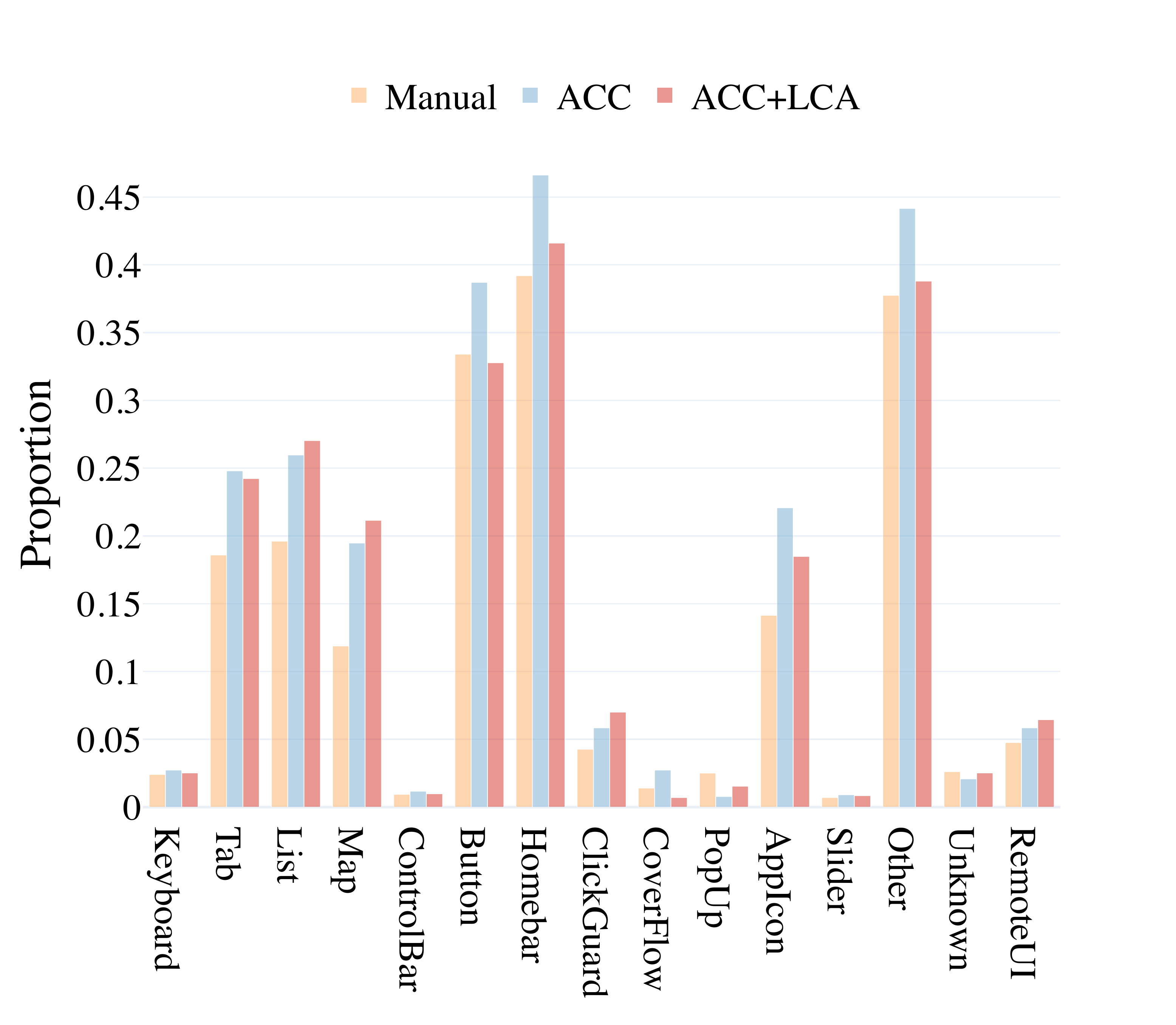}
    \caption{Proportion of sequences in which the driver interacted with the respective UI element}
    \label{fig:ElementBars}
    \end{subfigure}
    \qquad
    \begin{subfigure}{0.47\linewidth}
    \includegraphics[width = \linewidth]{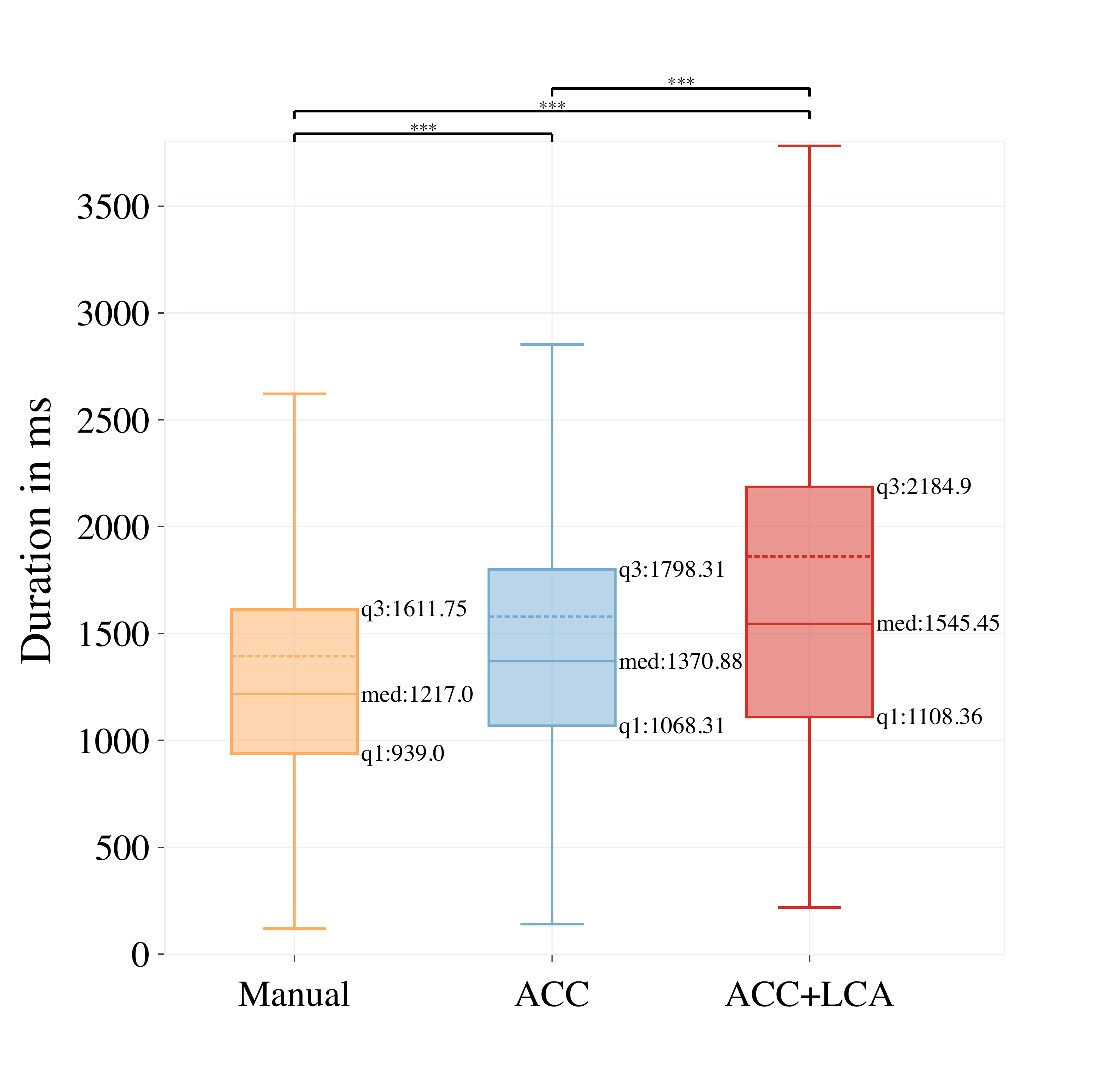}
    \caption{Mean glance duration toward the center stack touchscreen for different levels of driving automation. Significant differences are indicated as: $^{*}$p$<$0.1; $^{**}$p$<$0.05; $^{***}$p$<$0.01}
    \label{fig:Boxplots}
    \end{subfigure}
    \caption{The effect of the automation level on UI interactions and mean glance duration.}
\end{figure*}

Whereas the effect of ACC is significant across all models ($p<0.001$), the effect of ACC+LCA is only significant for \textit{Tab}, \textit{List}, \textit{Map}, and \textit{AppIcon}, but not for \textit{Homebar} and \textit{Button}. The odds that a driver performs a \textit{Map} or \textit{List} interaction when ACC and LCA are active are $e^{0.83} \approx 2.3$ and $e^{0.49} \approx 1.6$ times higher than the odds to perform these interactions in manual driving. These effects are also visualized in Figure~\ref{fig:ElementBars}.

The effect of 50--100 is significant for \textit{List}, \textit{Button}, \textit{Homebar}, and \textit{AppIcon} suggesting that drivers perform more touch interactions when driving between 50\,km/h and 100\,km/h compared to driving at speeds below 50\,km/h. The effect of 100+ is only significant for \textit{AppIcon}. The odds of drivers opening an application is $e^{0.25} \approx 1.3$ times higher at speeds above 100\,km/h compared to speeds between 0--50\,km/h. The effect of curved is significant in all models but the \textit{Homebar} model and suggests that drivers interact less with the UI in general during curved driving. The odds for a driver to interact with these elements in curved driving conditions are between $0.68$ and $0.87$ the odds compared to straight driving.

Across all models, our results suggest that the effect of driving automation on tactical self-regulation is larger than the effect of vehicle speed or road curvature. Furthermore the effect of driving automation is the largest for list and map interactions and the smallest for homebar and button interactions.

\subsection{Operational Self-Regulation}
Operational self-regulation is evaluated by identifying how drivers adapt their glance behavior in terms of mean glance duration and long glance probability. The results of our (generalized) linear mixed-effect models (see Table~\ref{tab:GlanceResults}) suggest that drivers adapt their glance behavior while interacting with the center stack touchscreen based on automation status, vehicle speed, and road curvature.

\begin{table*} \centering
\small
  \caption{Mixed-effects models for mean glance duration and long glance probability toward the center stack touchscreen. The coefficients and standard errors of the mean glance duration models are given on a logarithmic scale and the coefficients and standard errors for the long glance model represent log odds.} 
  \label{tab:GlanceResults} 
\begin{tabular}{@{\extracolsep{5pt}}lrrrr} 
\\[-1.8ex]\hline 
\hline \\[-1.8ex] 
 & \multicolumn{4}{c}{\textit{Dependent variable:}} \\ 
\cline{2-5} 
\\[-1.8ex] & \multicolumn{2}{c}{Mean Glance Duration} & \multicolumn{2}{c}{Long Glance} \\ 
\\[-1.8ex] & \multicolumn{2}{c}{\textit{linear}} & \multicolumn{2}{c}{\textit{generalized linear}} \\ 
 & \multicolumn{2}{c}{\textit{mixed-effects}} & \multicolumn{2}{c}{\textit{mixed-effects}} \\ 
\\[-1.8ex] & Model 1 & Model 2 & Model 3 & Model 4\\ 
\hline \\[-1.8ex] 
 Intercept & 7.13$^{***}$ (0.01) & 7.20$^{***}$ (0.01) & $-$0.94$^{***}$ (0.03) & $-$0.55$^{***}$ (0.06) \\ 
  ACC & 0.11$^{***}$ (0.02) & 0.03 (0.06) & 1.00$^{***}$ (0.09) & 0.45 (0.33) \\ 
  ACC+LCA & 0.19$^{***}$ (0.02) & 0.16$^{***}$ (0.04) & 1.29$^{***}$ (0.10) & 0.98$^{***}$ (0.23) \\ 
  50-100 &  & $-$0.07$^{***}$ (0.01) &  & $-$0.35$^{***}$ (0.07) \\ 
  100+ &  & $-$0.10$^{***}$ (0.02) &  & $-$0.55$^{***}$ (0.09) \\ 
  curved &  & $-$0.05$^{**}$ (0.02) &  & $-$0.44$^{***}$ (0.08) \\ 
  ACC:50-100 &  & 0.15$^{*}$ (0.07) &  & 0.78$^{*}$ (0.37) \\ 
  ACC+LCA:50-100 &  & 0.06 (0.05) &  & 0.56$^{*}$ (0.27) \\ 
  ACC:100+ &  & 0.10 (0.07) &  & 0.83$^{*}$ (0.36) \\ 
  ACC+LCA:100+ &  & 0.005 (0.06) &  & 0.19 (0.30) \\ 
  ACC:curved &  & 0.07 (0.11) &  & 0.46 (0.55) \\ 
  ACC+LCA:curved &  & 0.07 (0.10) &  & $-$0.53 (0.51) \\ 
  50-100:curved &  & $-$0.08$^{***}$ (0.02) &  & $-$0.26 (0.13) \\ 
  100+:curved &  & $-$0.04 (0.04) &  & $-$0.73$^{*}$ (0.29) \\ 
  ACC:50-100:curved &  & $-$0.13 (0.12) &  & $-$1.07 (0.64) \\ 
  ACC+LCA:50-100:curved &  & $-$0.04 (0.12) &  & 0.29 (0.61) \\ 
  ACC:100+:curved &  & $-$0.22 (0.14) &  & $-$0.53 (0.79) \\ 
  ACC+LCA:100+:curved &  & 0.13 (0.14) &  & 2.25$^{**}$ (0.79) \\ 
 \hline \\[-1.8ex] 
Akaike Inf. Crit. & 12,640.22 & 12,582.53 & 12,288.55 & 12,144.72 \\ 
Bayesian Inf. Crit. & 12,676.34 & 12,727.01 & 12,317.45 & 12,281.98 \\ 
\hline 
\hline \\[-1.8ex] 
\textit{Note:}  & \multicolumn{4}{r}{$^{*}$p$<$0.05; $^{**}$p$<$0.01; $^{***}$p$<$0.001} \\ 
\end{tabular} 
\end{table*} 

\subsubsection{Mean Glance Duration}\label{ch:resultsAverageGlance} The results of Model 1 as shown in Table~\ref{tab:GlanceResults} suggest that the effect of ACC and ACC+LCA on drivers' mean glance duration toward the center stack touchscreen is significant ($p<0.001$) compared to manual driving. As the mean glance duration is measured on a logarithmic scale, the exponent of models’ coefficients can be interpreted roughly as percent changes. When ACC is active, drivers' mean glance duration increases by $e^{0.11} \approx 1.12 = 12\,\%$. When ACC and LCA are both active, drivers' mean glance duration increases by $20\,\%$ compared to manual driving. Post-hoc testing reveals that the difference between ACC and ACC+LCA is also significant. The effects are shown in Figure~\ref{fig:Boxplots}.


In addition to Model 1, Model 2 adds vehicle speed, road curvature, and the accompanying interactions as fixed effects. In this model the combination of manual and straight driving, at speeds between $0-50\,km/h$ serves as a reference and all coefficients displayed in Table~\ref{tab:GlanceResults} need to be interpreted accordingly. Apart from the significant main effects for ACC+LCA, 50-100, 100+, and curved, the interaction between ACC and 50-100 is also significant. 

\begin{figure*}
	\centering
	\includegraphics[width=\linewidth]{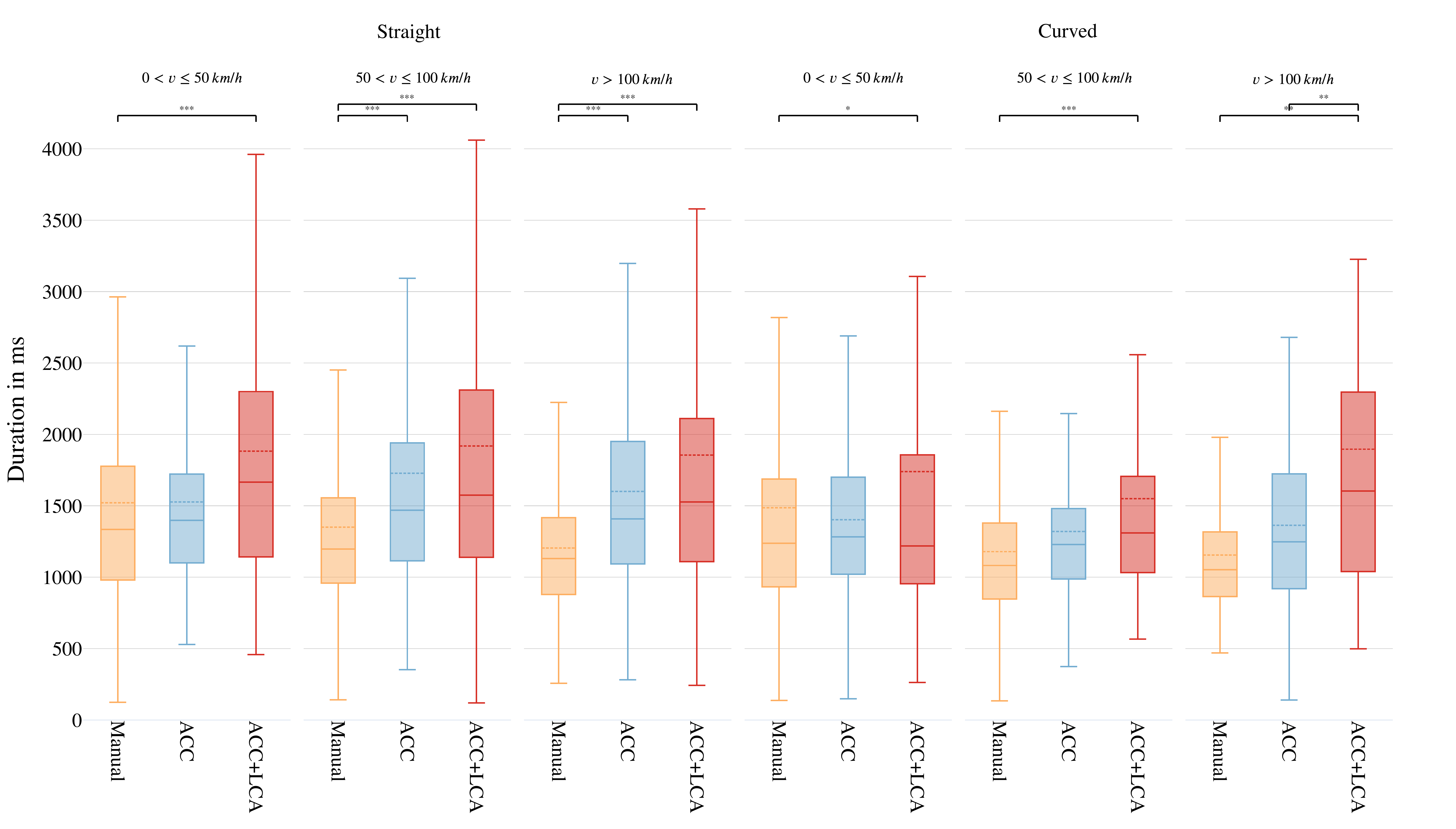} 
	\caption{Boxplots of the mean glance duration toward the center stack touchscreen grouped according to road curvature (left and right half), vehicle speed (combination of three boxplots each), and automation status (by color). Statistical significant results of Tukey's pairwise post-hoc test are indicated as: $^{*}$p$<$0.1; $^{**}$p$<$0.05; $^{***}$p$<$0.01} 
	\label{fig:BoxplotsAll} 
\end{figure*}

Furthermore, we are interested in whether the effect of ACC and ACC+LCA driving on drivers' self-regulation differs depending on the driving situations. We, therefore, perform pairwise post-hoc comparisons as shown in Figure~\ref{fig:BoxplotsAll}. We adjust p-values based on Tukey's method for comparing a family of three estimates.

For straight driving at speeds between 0 and 50\,km/h there is a significant difference in mean glance duration between manual driving and ACC+LCA driving ($p = 0.0008$). During ACC+LCA driving, drivers' mean glance duration increases by 17\,\%. For vehicle speeds between 50 and 100\,km/h and for speeds over 100\,km/h we observe significant differences for both, ACC and ACC+LCA, compared to manual driving ($p < 0.0001$ for all conditions). During ACC+LCA driving, drivers increase their mean glance duration by 26\,\% (50-100\,km/h) and 17\,\% (100+\,km/h) respectively. Whereas Figure~\ref{fig:BoxplotsAll} also indicates differences between the ACC and ACC+LCA conditions, they are not significant.  

During curved driving, we find significant differences between manual and ACC+LCA across all vehicle speeds, but no significant difference between manual and ACC. Furthermore, the difference between ACC driving and ACC+LCA driving at speeds above 100\,km/h is significant with a 47\,\% increase in the mean glance duration during ACC+LCA driving.

\subsubsection{Long Glance Probability}
The results of Model 3, as presented in Table~\ref{tab:GlanceResults}, suggest that the effect of ACC and ACC+LCA driving on the probability that a driver performs a long glance during an interaction sequence is significant. The odds that a driver performs a long glance toward the center stack touchscreen are $e^{1} \approx 2.7$ (ACC) and $e^{1.29} \approx 3.6$ (ACC+LCA) times higher compared to manual driving. Post-hoc pairwise comparisons also reveal a significant difference between ACC and ACC+LCA with the odds being 1.3 times higher ($p=0.0495$) in the ACC+LCA condition. 

\begin{figure*}
	\centering
	\includegraphics[width=\linewidth]{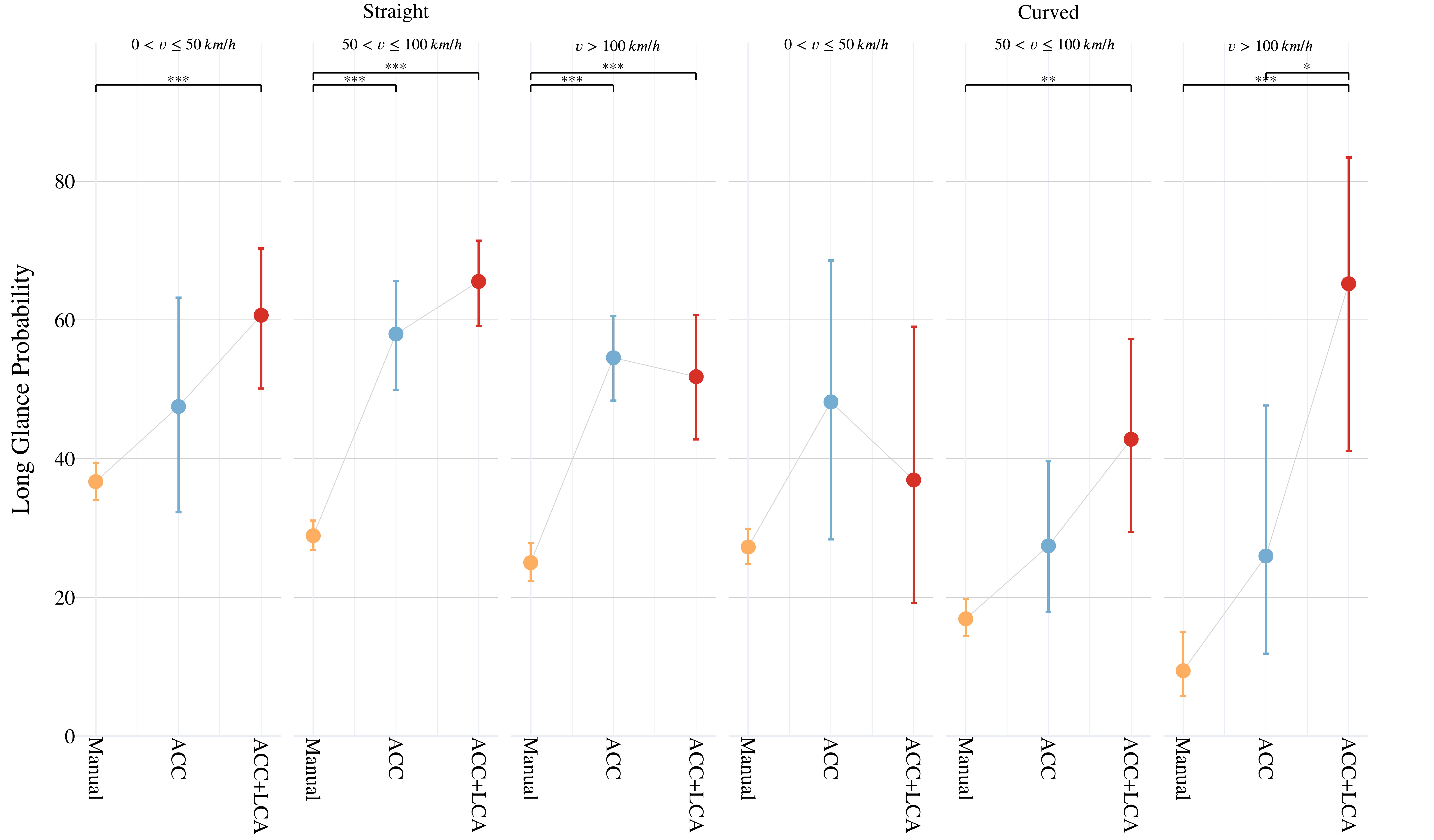} 
	\caption{Marginplot of the predicted long glance probabilities and accompanying confidence intervals. Significant results according to Tukey's pairwise post-hoc test are indicated as: $^{*}$p$<$0.1; $^{**}$p$<$0.05; $^{***}$p$<$0.01} 
	\label{fig:LongGlancePredictions} 
\end{figure*}

The results of Model 4 show significant effects of vehicle speed, road curvature, and various interactions. Comparing the main effects, we observe that compared to the reference the effect of ACC+LCA is almost double the effects of 50-100, 100+, or curved.
The model predictions and confidence intervals are visualized in Fig~\ref{fig:LongGlancePredictions}. Post-hoc tests were performed using Tukey's multiple comparison method.

For interactions, while straight driving with 0-50\,km/h we observe a significant difference in the long glance probability between manual driving and ACC+LCA driving. Whereas the results also suggest a difference between manual driving and ACC driving, the difference is not significant. However, for vehicle speeds between 50-100\,km/h and speeds above 100\,km/h, the long glance probability is significantly higher in ACC and ACC+LCA driving compared to manual driving. For straight driving, no significant difference in the long glance probability between ACC driving and ACC+LCA driving can be observed. For curved driving at speeds of  0-50\,km/h no significant differences between manual, ACC, and ACC+LCA driving are observed (see Figure~\ref{fig:LongGlancePredictions}). For speeds of 50-100\,km/h a significant difference between manual driving and ACC+LCA is shown. For vehicle speeds above 100\,km/h the difference between manual and ACC+LCA driving is significant and the difference between ACC and ACC+LCA driving is also significant. However, for curved driving, no significant difference can be observed between manual driving and ACC driving, even though the effect sizes of the pairwise comparison suggest so.

\section{Discussion}

\subsection{The Effect of Driving Automation on Tactical Self-Regulation}
Our findings on drivers' tactical self-regulation suggest that drivers adapt their interactions with the center stack touchscreen-based on the automation level, vehicle speed, and road curvature (RQ1). Unlike \citeauthor{noble.2021}~\cite{noble.2021}, we observed that the automation level significantly influences the probability to engage in a secondary task. 
During automated driving, drivers interact particularly more often with lists or maps compared to the homebar or general buttons. A potential explanation for this behavior is that lists and maps are visually more complex and drivers seem to perform these interactions in less demanding driving situations, e.g., with automation enabled or while driving straight. In contrast, the homebar, for example, is easy to access as it is visible on every screen and always located in the same position. Therefore, drivers' engagement with it is not much affected by the driving demand. These findings are in line with previous work~\cite{schneidereit.2017, morgenstern.2020, onate-vega.2020, oviedo-trespalacios.2018,choudhary.2017}. However, we are the first to use naturalistic driving data to measure tactical self-regulation on the level of specific UI interactions.

\subsection{The Effect of Driving Automation on Operational Self-Regulation}
Our results provide novel insights into drivers' operational self-regulation. We show that drivers not only adapt their glance behavior based on automation level (RQ2), vehicle speed, and road curvature but show that significant interdependencies between these factors exist (RQ3). Drivers extend the margins to which they consider it safe to focus on the center stack touchscreen with an increasing level of driving automation, even though they are supposed to constantly supervise the driving automation~\cite{SAE.2021}. The mean glance duration in ACC+LCA driving is 0.45\,s longer than in manual driving. This is in line with \citeauthor{morando.2021}~\cite{morando.2021}, who report an average increase of 0.3\,s. In line with the findings of \citeauthor{noble.2021}~\cite{noble.2021}, \citeauthor{gaspar.2019}~\cite{gaspar.2019}, and \citeauthor{morando.2021}~\cite{morando.2021}, we also show that drivers are more likely to perform glances longer than two seconds when driving automation is enabled. Whereas \citeauthor{morando.2021}~\cite{morando.2021} report an increase in the long glance probability toward the center stack touchscreen between manual and level 2 driving of 425\,\%, our results suggest an increase of 263\,\%. While the trend is similar, the absolute difference is probably due to different data acquisition.

We also show significant differences in the mean glance duration between manual, ACC, and ACC+LCA driving. Whereas the difference between ACC and LCA+ACC for straight driving is rather small, it becomes more prominent in curved driving over 50\,km/h. This indicates that drivers are aware of the different benefits of the respective functions and seem to trust its functionality.
Whereas \citeauthor{noble.2021}~\cite{noble.2021} and \citeauthor{morando.2019}~\cite{morando.2019} found no significant differences in the average off-road glance duration for ACC or LCA driving compared to manual driving, we show significant differences across all conditions. There may be two reasons for this: First, the amount of data we considered in our study is larger. Second, our eye tracker explicitly detects glances toward the center stack touchscreen that we then map to UI interactions. In other studies~\cite{morando.2019, risteska.2021, noble.2021, yang.2021}, authors could not differentiate between general off-path glances, which might still be driving-related, and distraction-related off-path glances. This, inevitably, increases the number of false positives, making it harder to obtain significant results.

\subsection{Limitations}

Naturalistic driving studies allow us to observe drivers in their natural driving environment. Driving simulator studies or test track studies, in contrast, suffer from an \emph{instruction effect} because participants need to perform specific predefined tasks~\cite{carsten.2017}. Furthermore, by leveraging production systems, we collect a large amount of data without the need for, potentially, error-prone manual labeling. However, certain limitations should be considered when interpreting the results.
The cars that contributed to the data collection are company internal test cars. They are used for various testing procedures but also for transfer and leisure rides of employees. However, we did not observe any indications in the data that specific UI stress tests are performed while driving. Furthermore, even when drivers follow a certain test protocol to evaluate driving-related functions, we argue that the incentive to interact with the \acp{IVIS} does not differ from real-world driver behavior.  
Furthermore, drivers need to be considered expert users as they are familiar with the cars and additionally obtained a prototype driver's license. However, the effect this might have is not entirely clear. Whereas more experienced drivers tend to distribute their visual attention more adequately~\cite{wikman.1998}, \citeauthor{naujoks.2016} report that drivers who are familiar with driving assistance systems are more likely to engage in secondary tasks during assisted driving compared to drivers with no experience~\cite{naujoks.2016}. In general, the glance duration distribution is roughly similar to those reported in related studies~\cite{gaspar.2019, morando.2019, noble.2021}.
Due to data privacy regulations, we cannot differentiate between individual drivers. Considering that more than 100 cars, that are driven by even more individual drivers, contributed to the data collection, the risk of overfitting to particular drivers is small. However, it is important to consider that only employees contributed to the data collection. For this reason, the results are likely to be biased toward mid-age drivers.
Finally, the sample size for some conditions is relatively small (30--50 sequences). This leads to non-significant differences even though visual inspection of the plots and model parameters suggest relatively large effects. More data could lead to improved results and current non-significant results should be interpreted with care as this does not imply that the effect does not exist.

\section{Conclusion}
We present the first naturalistic driving study to investigate tactical and operational self-regulation of driver interactions with center stack touchscreens. Understanding self-regulation is key to understanding the effects of automation and assistance functions on driver distraction and driving safety. 
The key strengths of our study over the state-of-the-art are two-fold: (1) The large amount of naturalistic data, compared to related approaches~\cite{morando.2019,noble.2021, naujoks.2016}, allows us to investigate drivers' tactical and operational self-regulation in greater detail concerning the driving demand. (2) We evaluate self-regulation specifically during interactions with the center stack touchscreen by combining driving data, UI interactions, and explicit glances toward the center stack touchscreen. That makes this the first naturalistic driving study to show self-regulation based on the analysis of touchscreen interactions.

Our modeling results show that driving automation has a stronger effect on self-regulation than vehicle speed or road curvature. Drivers interact more with the \acs{IVIS} when ACC or ACC+LCA is enabled and use complex UI elements such as lists and maps twice as often when driving assistance is active. Even though driving assistance functions up to level 2 still demand the driver to have full control over the car, we observe 20\% longer off-road glances when ACC+LCA is active. 

Further research is needed, but based on the assumption that drivers kept the driving similarly safe throughout all conditions, fixed limits for acceptable demand as reported in the NHTSA Driver Distraction Guidelines~\cite{NHTSA.2014} need to be adjusted according to different levels of driving automation and driving demands.

%

\balance
\bibliographystyle{ACM-Reference-Format}
\bibliography{self-regulation.bib}

\end{document}